\documentstyle[12pt,a41,epsfig,bezier]{article}

\def\photonatomup{\begin{picture}(1.5,3)(0,0)
                             \put(-0.75,3){\tencircw \symbol{3}}
                             \put(-0.75,1.5){\tencircw \symbol{2}}
                             \put(0.75,1.5){\tencircw \symbol{0}}
                             \put(0.75,0){\tencircw \symbol{1}}
                   \end{picture}
                  }

\def\photonup{\begin{picture}(1.5,30)(0,0)
                  \multiput(0,0)(0,3){10}{\photonatomup}
               \end{picture}
              }

\def\fermionurr{\begin{picture}(30,15)(0,0)
                        \put(-30,-15){\vector(2,1){15}}
                        \put(-15,-7.5){\line(2,1){15}}
                  \end{picture}
                 }

\def\fermiondrr{\begin{picture}(30,15)(0,0)
                        \put(0,0){\vector(2,-1){15}}
                        \put(15,-7.5){\line(2,-1){15}}
                  \end{picture}
                 }

\newenvironment{Feynman}[3]{\begin{center}
                            \setlength{\unitlength}{#3 mm}
                            \begin{picture}(#1)(#2)
                            \thicklines
                           }{\end{picture} \end{center}}

\newcommand{\nll}{\nonumber \\}

\newcommand{\bq}{\begin{equation}}
\newcommand{\eq}{\end{equation}}
\newcommand{\ba}{\begin{eqnarray}}
\newcommand{\ea}{\end{eqnarray}}

\newcommand\GeV{\,\mbox{GeV}}

\newcommand{\ds}{\displaystyle}

\newcommand{\nnn}{\nonumber}

\newcommand {\Mp }{\mbox{$M^2        $}}

\newcommand {\Qe  }{\mbox{$Q_{e}  $}}
\newcommand {\COSPHI }{\mbox{$\cos{\phi} $}}
\newcommand {\pspin}{{\it s}}
\newcommand {\Pe   }{\mbox{$p_e$}}
\newcommand {\pe   }{\mbox{$p_e$}}

\begin{document}
\begin{flushleft} 
{\tt hep-ph/9711228} \\    {October 1997}
\end{flushleft}

\begin{center}
{\LARGE\bf ${\cal{O}}(\alpha)$ QED Corrections to Polarized \\
Elastic $\mu e$ and Deep Inelastic $l N$ Scattering}

\vspace{1cm}

{Dima Bardin$^{a,b,c}$,~Johannes Bl\"umlein$^a$,~Penka Christova$^{a,d}$,
and Lida Kalinovskaya$^{a,c}$}

\vspace*{1.0cm}

{\it $^a$ DESY--Zeuthen,}
{\it Platanenallee 6, D--15735 Zeuthen, Germany }

\vspace{1mm}
{\it $^b$ INFN, Sezione di Torino, Torino, Italy}

\vspace{1mm}
{\it $^c$ JINR, ul. Joliot-Curie 6, RU--141980 Dubna, Russia}

\vspace{1mm}
{\it $^d$ Bishop Konstantin Preslavsky University of Shoumen,}
{\it 9700 Shoumen, Bulgaria   } \\

\vspace*{2cm}

\end{center}

\begin{abstract}
\noindent
Two computer codes relevant for the description of deep inelastic
scattering off polarized targets are discussed. The code
${\mu}${\bf e}{\it la} deals with radiative corrections to elastic
$\mu e$ scattering, one method applied for muon beam polarimetry.
The code {\tt HECTOR} allows to calculate both the radiative corrections
for unpolarized and polarized deep inelastic scattering, including
higher order QED corrections.
\end{abstract}

\section{Introduction}

\vspace{1mm}
\noindent
The exact knowledge of QED, QCD, and electroweak (EW) radiative
corrections (RC) to the deep inelastic scattering (DIS) processes
is necessary for a precise determination of the nucleon structure
functions.
The present and forthcoming high statistics measurements of polarized
structure functions
in the SLAC experiments, by HERMES, and later by COMPASS require
the knowledge of the RC to the DIS polarized cross-sections at the
percent level.

Several codes based on different approaches
for the calculation of the RC to DIS experiments, mainly for
non-polarized DIS, were developped and thoroughly compared
in the past, cf.~\cite{h91}.
Later on the radiative corrections for a vast amount of experimentally
relevant sets of kinematic variables were calculated~\cite{JBKIN},
including also semi-inclusive situations as the RC's in the case of
tagged photons~\cite{tp}. Furthermore the radiative corrections
to elastic $\mu$-$e$ scattering, a process to monitor (polarized)
muon beams, were calculated~\cite{muela}. The corresponding codes are~:

\begin{itemize}

\item {\tt HECTOR 1.00}, (1994-1995)~\cite{he},
by the Dubna-Zeuthen Group.
It calculates QED, QCD and EW corrections for variety of measuremets for
unpolarized DIS.

\item ${\mu}${\bf e}{\it la}~{\tt 1.00}, (March 1996)~\cite{muela},
calculates ${\cal{O}}(\alpha)$ QED correction for
polarized ${\mu}e$ elastic scattering.

\item {\tt HECTOR 1.11}, (1996)
extends {\tt HECTOR 1.00} including the radiative corrections for
polarized DIS~\cite{ph}, and for DIS with tagged photons~\cite{tp}.
The beta-version of the code is available from {\tt http://www.ifh.de/}.
\end{itemize}

\section{The Program ${\mu}${\bf e}{\it la} }

\vspace{1mm}
\noindent
Muon beams may be monitored using the processes of $\mu$ decay and
$\mu e$ scattering in case of atomic targets. Both processes were
used by the SMC experiment. Similar techniques will be used by the
COMPASS experiment. For the cross section measurement
the radiative corrections to these processes have to be known at high
precision. For this purpose a renewed calculation of the radiative
corrections to $\sigma(\mu e \rightarrow \mu e)$
was performed~\cite{muela}.

The differential cross-section of polarized elastic $\mu e$ scattering
in the Born approximation reads, cf.~\cite{SMC},
\ba
\frac{d\sigma^{^{\rm{BORN}}}}{dy} = \frac{2\pi\alpha^2}{m_eE_{\mu}}
     \left[\frac{(Y-y)}{y^2Y}\left(1-yP_{e}P_{\mu}\right)
     + \frac{1}{2}\left(1-P_{e}P_{\mu} \right)\right],
\ea
where $\ds{y=y_{\mu}=1-{E^{'}_{\mu}}/{E_{\mu}}=
{E^{'}_{e}}/{E_{\mu}}=y_{e}}$,
$\ds{Y=\left(1+{m_{\mu}}/2/E_{\mu}\right)^{-1}} = y_{max}$,
$m_{\mu},\;m_e$ -- muon and electron masses,
$E_{\mu},\;E^{'}_{\mu},\;E^{'}_{e}$ the energies of the incoming and
outgoing muon, and outgoing electron respectively, in the laboratory
frame.
$P_{\mu}$ and $P_{e}$ denote the longitudinal polarizations
of muon beam and electron target. At Born level
$y_{\mu}$ and $y_e$ agree. However, both quantities are different
under inclusion of radiative corrections due to bremsstrahlung.
The correction factors may be rather different depending on which
variables ($y_{\mu}$ or $y_e$) are used.

In the SMC analysis the $y_{\mu}$-distribution was used to measure
the electron spin-flip asymmetry $A^{\rm{exp}}_{\mu e}$.
Since previous  calculations, \cite{blkwasborn,bla},
referred to $y_e$, and only ref.~\cite{bla} took polarizations
into account, a new calculation was performed, including
the complete ${\cal{O}}(\alpha)$ QED correction for the
$y_{\mu}$-distribution, longitudinal polarizations for both leptons,
the $\mu$-mass effects, and neglecting $m_e$ wherever possible.
Furthermore the present calculation allows for cuts on the
electron recoil energy ($35 \GeV$),
the energy balance ($40 \GeV$), and angular cuts for both
outgoing leptons (1~mrad).
The default values are given in parentheses.

Up to order ${\cal{O}}(\alpha^3)$,
14 Feynman graphs contribute to the cross-section for $\mu$-$e$
scattering, which may be subdivided into
$12 = 2 \times6$ pieces, which are separately gauge invariant
\ba
\frac{d\sigma^{^{\rm{QED}}}}{dy_{\mu}}=\sum_{l=1}^2\sum_{k=1}^6
\frac{d\sigma_k^l}{dy_{\mu}}~.
\label{cont12}
\ea
One may express (\ref{cont12}) also as
\begin{eqnarray}
\frac{d\sigma^{^{\rm{QED }}}}{dy_{\mu}}=\sum_k\left(
\frac{d\sigma_k^{\rm{unpol}}}{dy_{\mu}}+P_eP_{\mu}
\frac{d\sigma_k^{\rm{pol  }}}{dy_{\mu}}\right).
\end{eqnarray}
The indices $l$ and $k$ in the combinations $lk$ have the meaning
\[
\begin{array}{rccl}
l=&1& - & {\mbox{unpolarized contribution,}}\;l={\rm{unpol}};\\
  &2& - & {\mbox{polarized contribution (terms with}}\;P_eP_{\mu}\;
          {\mbox{in $d\sigma^{\rm{BORN}}$)}},\;l={\rm{pol}}. \\
\end{array}
\]
\[
\begin{array}{rccl}
k=&1&-& {\mbox{Born cross-section,}}\;k=b;\\
  &2&-& {\mbox{RC for the muonic current: vertex + bremsstrahlung,}}\;
        k={\mu\mu};\\
  &3&-& {\mbox{amm contribution from muonic current,}}\;k={\rm{amm}};\\
  &4&-& {\mbox{RC for the electronic current: vertex +
bremsstrahlung,}}\;
        k={ee};\\
  &5&-& \mu e\;{\mbox{interference: two-photon exchange +}}\\
  & & & {\mbox{muon-electron bremsstrahlung interference,}}\;
        k={\mu e};\\
  &6&-& {\mbox{vacuum polarization correction, running}}\;
\alpha,\;k={\rm{vp}}.
\end{array}
\]
The {\tt FORTRAN} code for the
scattering cross section (\ref{cont12})
$\mu${\bf e}{\it la} was used in a recent analysis of the SMC
collaboration.

The RC, $\delta^A_{y_{\mu}}$,
to the asymmetry $A^{^{\rm{QED}}}_{\mu e}$
shown in figures~1 and 2
is defined as
\ba
\delta^A_{y_{\mu}} = \frac{A^{^{\rm{QED }}}_{\mu e}}
{A^{^{\rm{BORN}}}_{\mu e}}-1\quad (\%),
\quad\mbox{where}\quad
A_{\mu e}={\displaystyle{\frac{d\sigma^{\rm{pol}}}{
d\sigma^{\rm{unpol}}}}}.
\ea
The results  may be summarized as follows. The ${\cal{O}}(\alpha)$ QED
RC to polarized elastic $\mu e$ scattering were  calculated for the
first time using the variable  $y_{\mu}$. A rather general {\tt FORTRAN}
code $\mu${\bf e}{\it la} for this process was created allowing for the
inclusion of kinematic cuts.
Since under the conditions of the SMC experiment the corrections turn
out to be small our calculation justifies their neglection.

\section{Program {\tt HECTOR}}
\subsection{Different approaches to RC for DIS}

\vspace{1mm}
\noindent
The radiative corrections to deep inelastic scattering are treated
using two basic approaches. One possibility consists in generating
events on the basis of matrix elements including the RC's. This approach
is suited for detector simulations, but requests a very hughe number of
events to obtain the corrections at a high precision. Alternatively,
semi-analytic codes allow a fast and very precise evaluation, even
including a series of basic cuts and flexible adjustment to specific
phase space requirements, which may be caused by the way kinematic
variables are experimentally measured, cf.~\cite{JBKIN,he}.
Recently, a third approach, the so-called deterministic approach,
was followed, cf.~\cite{da}.
It treats the RC's completely exclusively combining features of fast
computing with the possibility to apply any cuts.
Some elements of this approach were used in
${\mu}${\bf e}{\it la} and in the branch of {\tt HECTOR 1.11,} in  which
DIS with tagged photons is calculated.

Concerning the theoretical treatment three approaches are in use
to calculate the radiative corrections: 1) the model-independent
approach (MI);  2) the leading-log approximation (LLA); and 3) an
approach based on the quark-parton model (QPM) in evaluating the
radiative corrections to the scattering cross-section.

In the  model-independent  approach the QED corrections are only
evaluated for the leptonic tensor. Strictly it
applies only for neutral current processes. The hadronic tensor can be
dealt with in its most general form on the Lorentz-level.
Both lepton-hadron corrections as well as pure hadronic corrections
are neglected. This is justified in a series of cases in which these
corrections turn out to be very small.
The leading logarithmic approximation is one of the semi-analytic
treatments in which the different collinear singularities of
$O((\alpha \ln(Q^2/m_l^2))^n)$ are evaluated and other corrections are
neglected. The QPM-approach  deals with the full set of
diagrams on the quark level. Within this method, any corrections
(lepton-hadron interference, EW) can be included.
However, it has limited precision too, now due to use of QPM-model
itself. Details on the realization of these approaches within the code
{\tt HECTOR} are given in ref.~\cite{he,crad96}.

\subsection{$O(\alpha)$ QED Corrections for Polarized Deep Inelastic
Scattering}
\vspace{1mm}
To introduce basic notation, we show the Born diagram
\vspace*{.5cm}
\begin{center}
\begin{figure}[h]
\begin{minipage}[bthp]{178mm}{
\begin{center}
\begin{Feynman}{60,60}{0,0}{0.8}
%
\put(00,60){\fermiondrr}
\put(-11,62){$l^\mp(\vec k_1,m)$}  
\put( 50,62){$l^\mp(\vec k_2,m)$}
\put( 50,-07){$X(\vec p^{\;'},M_h)$}
\put(-09,-07){$p \, (\vec p  ,M)$}
\put(34,30){$\gamma, Z$}
\put(60,60){\fermionurr}
\put(30.5,15){\photonup}
\put(30,15){\circle*{5}}
\put(30,17){\line(2,-1){30}}
\put(30,13){\line(2,-1){28}}
\put(30,45){\circle*{1.5}}
\put(30,15){\fermionurr}
\put(30,15){\fermiondrr}
\end{Feynman}
\end{center}
}\end{minipage}
\vspace*{.25cm}
\end{figure}
\end{center}

\noindent and the Born cross-section,
which is  presented as the product of the leptonic and hadronic tensor
\bq
d\sigma_{{Born}}
 =  \frac {2\pi \alpha^2}{Q^4}
           y
 \Biggl[ L^{\mu\nu} W_{\mu\nu} \Biggr]
  dx dy,
\label{sborn2}
\eq
with
\bq
q=k_1-k_2, \qquad  Q^2 = -q^2 ,  \qquad  S = 2 (p.k_1),
\label{xyqs}
\eq
and the Bjorken scaling variables
\bq
  y = \frac{p.q}{p.k_1}, \qquad   x = \frac{Q^2}{S y}.
\label{lambdas}
\eq
For the hadronic tensor, we use the representation of
ref.~\cite{bk1}
\ba
 W_{\mu\nu}     &=& p^0 (2\pi)^6 \sum \int
          \langle p^{'}| {\cal J}_{\mu} |p     \rangle
          \langle p    | {\cal J}_{\nu} |p^{'} \rangle
          \delta^4 (\sum_{i} p^{'}_i- p^{'}) \prod_{i} d p^{'}_i
\nll\nll
  &=& \left(-g_{\mu\nu}+ \frac{q_\mu q_\nu }{q^2} \right)
                                              {\cal F }_1(x,Q^2)
 +\frac{\widehat{p_\mu}\widehat{p_\nu}}{p.q}
                                              {\cal F }_2(x,Q^2)
 - i e_{\mu\nu\lambda\sigma} \frac{q^\lambda p^\sigma}
                                { 2 p.q}  {\cal F }_3(x,Q^2)
\nll
  && + i e_{\mu\nu\lambda\sigma}\frac{q^\lambda {\pspin}^\sigma} {p.q}
                                              {\cal G }_1(x,Q^2)
     + i e_{\mu\nu\lambda\sigma}
          \frac {q^{\lambda}(p.q {\pspin}^\sigma
                      -{\pspin}.q p^\sigma)}{(p.q)^2}
                                            {\cal G }_2(x,Q^2)
\nll
  && +      \left[\frac{\widehat{p_\mu} \widehat{{\pspin}_\nu}
        +   \widehat {{\pspin}_\mu} \widehat{p_\nu}}{2}
                -  {\pspin}.q
            \frac{\widehat{p_\mu}\widehat{ p_\nu}}{p.q}\right]
                     \frac{1}{p.q}                {\cal G }_3(x,Q^2)
\nll
  && +   {\pspin}.q \frac{\widehat{p_\mu} \widehat{p_\nu}}{(p.q)^2}
                                                  {\cal G }_4(x,Q^2)
               +  \left( - g_{\mu\nu}+ \frac{q_\mu q_\nu }{q^2}\right)
                  \frac{ {\pspin}.q  }{p.q}       {\cal G }_5(x,Q^2),
\label{hadten}
\ea
where
\ba
 \widehat {p_\mu}   =   p_\mu - \frac{p.q }{q^2}  q_\mu,\qquad
 \widehat {{\pspin}_\mu}  =  {\pspin}_\mu
- \frac{{\pspin}.q}{q^2}  q_\mu,
\nnn
\ea
and $\pspin$ is the four vector of nucleon polarization, which is given
by $s= \lambda_p M (0,{\vec n} )$
in the nucleon rest frame.

The combined  structure functions in eq.~(\ref{hadten})
\ba
{\cal  F }_{1,2}(x,Q^2)
 &=& Q^2_{e} F^{\gamma\gamma}_{1,2}(x,Q^2) + 2 |\Qe|
 \left( v_l -\pe \lambda_l a_l \right)\chi(Q^2)
F^{\gamma Z}_{1,2}(x,Q^2) \nll
      &&+
 \left( v^2_l + a^2_l - 2\pe\lambda_l v_l a_l
\right) \chi^2(Q^2)
 F^{ZZ}_{1,2}(x,Q^2),\nll
{\cal F }_{3}(x,Q^2)   &=&
 2|\Qe|\left(\pe a_l -\lambda_l v_l \right) \chi(Q^2)
F^{\gamma Z}_3(x,Q^2),  \nll
      &&+ \left[2\pe v_l a_l -\lambda_l
          \left( v^2_l + a^2_l\right)\right] \chi^2(Q^2)
F^{ZZ}_3(x,Q^2),    \nll
{\cal G }_{1,2}(x,Q^2) &=&
 -Q^2_{e} \lambda_l  g^{\gamma \gamma}_{1,2}(x,Q^2)
 + 2 |\Qe| \left(\pe a_l -\lambda_l v_l \right)\chi(Q^2)
g^{\gamma Z}_{1,2}(x,Q^2),   \nll
  && +  \left[2\pe v_l a_l - \lambda_l
         \left( v^2_l + a^2_l \right)\right] \chi^2(Q^2)
g^{ZZ}_{1,2}(x,Q^2),       \nll
{\cal G }_{3,4,5}(x,Q^2) &=&
              2 |\Qe|\left( v_l -\pe \lambda_l a_l \right) \chi(Q^2)
g^{\gamma Z}_{3,4,5}(x,Q^2),   \nll
  && +  \left( v^2_l + a^2_l - 2\pe\lambda_l v_l a_l \right)
                                         \chi^2(Q^2)
                                        g^{ZZ}_{3,4,5}(x,Q^2),
\ea
are expressed via the hadronic structure functions, the
$Z$-boson-lepton couplings  $v_l,\;a_l$,
and the ratio of the propagators for the photon and $Z$-boson
\ba
 \chi (Q^2) =
{G_\mu \over\sqrt{2}}{M_{Z}^{2} \over{8\pi\alpha}}{Q^2 \over
{Q^2+M_{Z}^{2}}}.
\ea
\noindent Furthermore we use the parameter $\Pe$ for which
$\Pe = 1$   for a scattered lepton and
$\Pe =-1$   for a scattered antilepton.
The hadronic structure functions
can be expressed in terms of parton densities accounting for the
twist-2 contributions only, see~\cite{bk1}. Here, a series of relations
between the different structure functions are used in leading order QCD.

The DIS cross-section on the Born-level
\ba
\frac{d^2\sigma_{\rm{Born}}}{dxdy}=
\frac{d^2\sigma^{\rm{unpol}}_{\rm{Born}}}{dxdy}+
\frac{d^2\sigma^{\rm{pol}}_{\rm{Born}}}{dxdy},
\ea
contains two contributions, the unpolarized part
\ba
\frac{\ds d\sigma^{   {unpol}}_{   {Born}}}{\ds dx dy}
 =    \frac{2\pi \alpha^2 }{Q^4}S
\sum_{i=1}^{3} \;  S^{U}_{i}(x,y) {\cal F}_i(x,Q^2),
\ea
with the kinematic functions
\ba
  S^{U}_{1}(y,Q^2) &=&  2xy^2,                            \nll
  S^{U}_{2}(y,Q^2) &=&   2 \left[(1-y)-\frac{xy \Mp}{S}\right],     \nll
  S^{U}_{3}(y,Q^2) &=&  x\left[1-(1-y)^2\right],
\label{unpols}
\ea
and the polarized part
\ba
\frac{\ds d\sigma^{   {pol}}_{   {Born}}}{\ds d x dy}
&=& \frac{2 \pi \alpha^2}{Q^4}\lambda^p_{N}f^p S
     \sum_{i=1}^{5} \;  S^{p}_{gi}(x,y) {\cal G}_i(x,Q^2).
\ea
Here,
$S^{p}_{gi}(x,y)$ are
functions, similar to~(\ref{unpols}),
and  may be found in~\cite{ph}.
Furthermore we used the abbrevations
\ba
f^{^{L}}&=&1, \qquad
\vec{n}^{^{L }}\;=\;\lambda^p_{N}\frac{\vec{k}_1}{|\vec{k}_1|},
\nll
f^{^{T}}&=&
      \COSPHI \; \frac{d\phi}{2 \pi}
     \sqrt{\frac{4\Mp x}{Sy}\left(1-y-\frac{\Mp xy}{S}\right)}
\;=\;
     \COSPHI \; \frac{d\phi}{2 \pi}
       \frac{1-y}{y}\sin\theta_2,
\nll
\vec{n}^{^{T}}&=&\lambda^{^{T}}_{N}{\vec{n}}_{\perp},
\qquad \mbox{with $\vec{k_1}.{\vec{n}}_{\perp}=0$}.
\ea

\noindent
The $O(\alpha)$ DIS cross-section reads
\ba
\frac{d^2 \sigma_{\rm QED, 1}}{d x_l d  y_l} =
\frac{\alpha}{\pi} \delta_{\rm VR}
\frac{d^2 \sigma_{\rm Born}}{d x_l d y_l} +
\frac{d^2 \sigma_{\rm Brems}}{d x_l dy_l} =
\frac{d^2 \sigma_{\rm QED,1}^{\rm unpol}}{d x_l d y_l} +
\frac{d^2 \sigma_{\rm QED,1}^{\rm pol}}{d x_l dy_l}.
\ea
All partial  cross-sections have a form similar
to the Born cross-section and are expressed
in terms of kinematic functions and combinations of structure functions.
In the $O(\alpha)$ approximation
the measured  cross-section, $\sigma_{\rm rad}$, is define as
\ba
\frac{d^2 \sigma_{\rm rad}}{d x_l d  y_l} &=&
\frac{d^2 \sigma_{\rm Born}}{d x_l d  y_l} +
\frac{d^2 \sigma_{\rm QED,1}}{d x_l d  y_l}
\;=\;
\frac{d^2 \sigma_{\rm rad}^{\rm unpol}}{d x_l d y_l} +
\frac{d^2 \sigma_{\rm rad}^{\rm pol}}{d x_l dy_l}.
\ea
In the
four following figures we illustrate the  RC-factor
\ba
\delta=\frac{d^2 \sigma_{\rm rad}}{d^2 \sigma_{\rm Born}}-1.
\ea
The radiative corrections calculated for leptonic variables grow towards
high $y$ and smaller values of $x$. The figures compare the results
obtained in LLA, accounting for initial ($i$) and final state ($f$)
radiation, as well as the Compton contribution ($c2$) with the result
of the complete calculation of the leptonic corrections. In most of
the phase space the LLA correction provides an excellent description,
except of extreme kinematic ranges.

A comparison of the radiative corrections for polarized deep inelastic
scattering between the codes {\tt HECTOR} and {\tt POLRAD}~\cite{pr}
was carried out. It had to be performed under {\it simplified}
conditions due to the restrictions of {\tt POLRAD}.
Corresponding results may be found in~\cite{crad96,hy,db}.

\subsection{Conclusions}

\noindent
For the evaluation of the QED radiative corrections to deep inelastic
scattering of polarized targets two codes {\tt HECTOR} and {\tt POLRAD}
exist. The code {\tt HECTOR} allows a completely general study of the
radiative corrections
in the model independent approach in $O(\alpha)$ for neutral
current reactions including $Z$-boson exchange.
Furthermore, the LLA corrections are available
in 1st and 2nd order, including soft-photon resummation and for
charged current reactions.
{\tt POLRAD} contains a branch which
may be used for some
semi-inclusive DIS processes. The initial state radiative corrections
(to 2nd order in LLA + soft photon exponentiation)
to these (and many more processes)  can be calculated in detail
with the code
{\tt HECTOR}, if the corresponding user-supplied routine {\tt USRBRN}
is used together with this package. This applies both for neutral
and charged current processes as well as a large variety of different
measurements of kinematic variables.
Aside the leptonic corrections, which were studied in detail already,
further investigations may concern QED corrections to the hadronic
tensor as well as the interference terms.


\newpage

\vspace*{1.0cm}
\hspace*{0.5cm}
\begin{figure}[htbp]
\begin{center}
\centerline{\epsfig{file=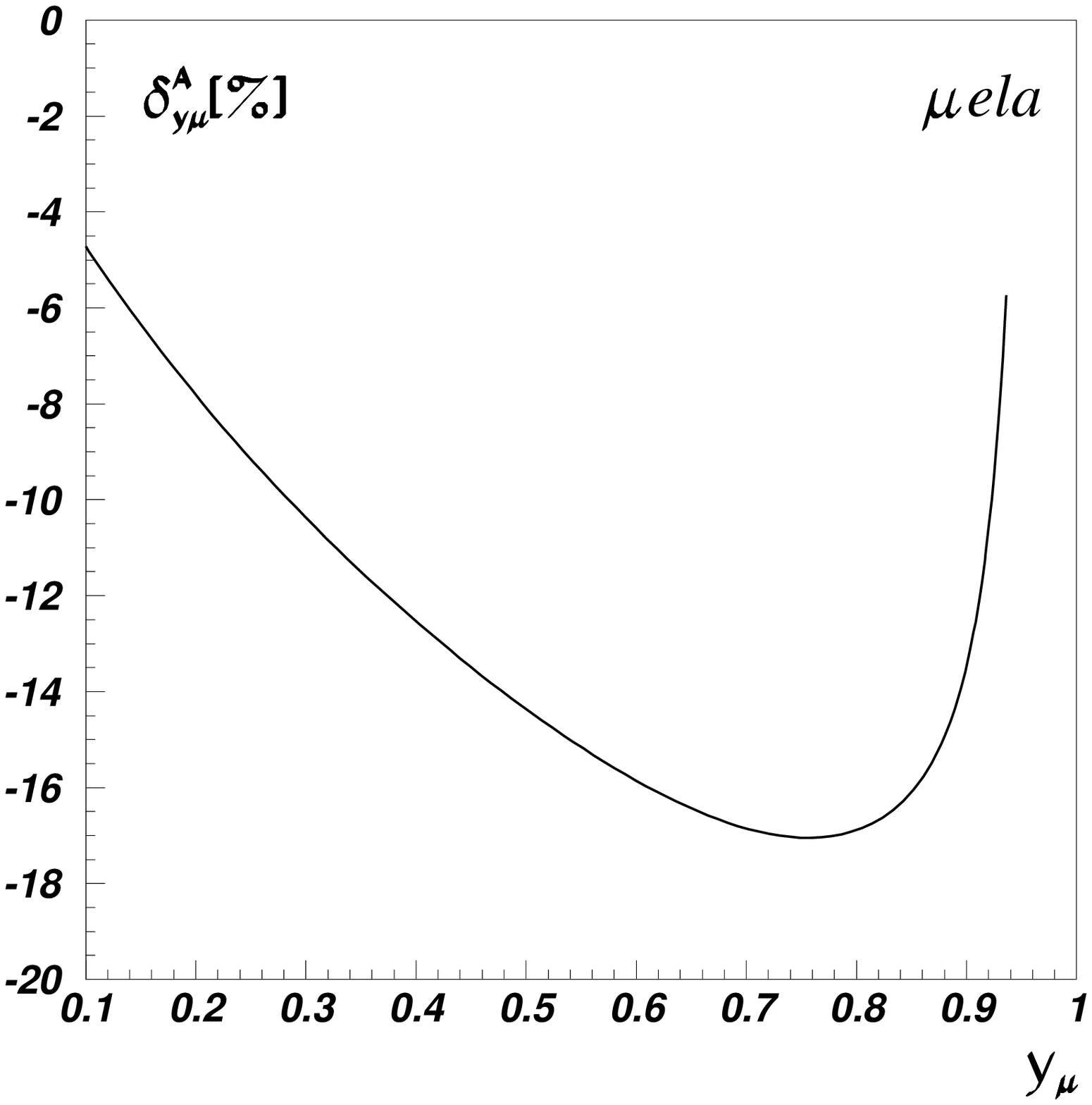,bbllx=0pt,bblly=0pt,%
bburx=600pt,bbury=600pt,width=10cm,angle=0}}
\end{center}
\vspace{-4.0cm}
\caption[xxx]{\sf The QED radiative corrections to asymmetry without
experimental cuts}.
\label{fiia1}
\end{figure}

\vspace*{0.5cm}
\hspace*{0.5cm}
\begin{figure}[htbp]
\begin{center}
\centerline{\epsfig{file=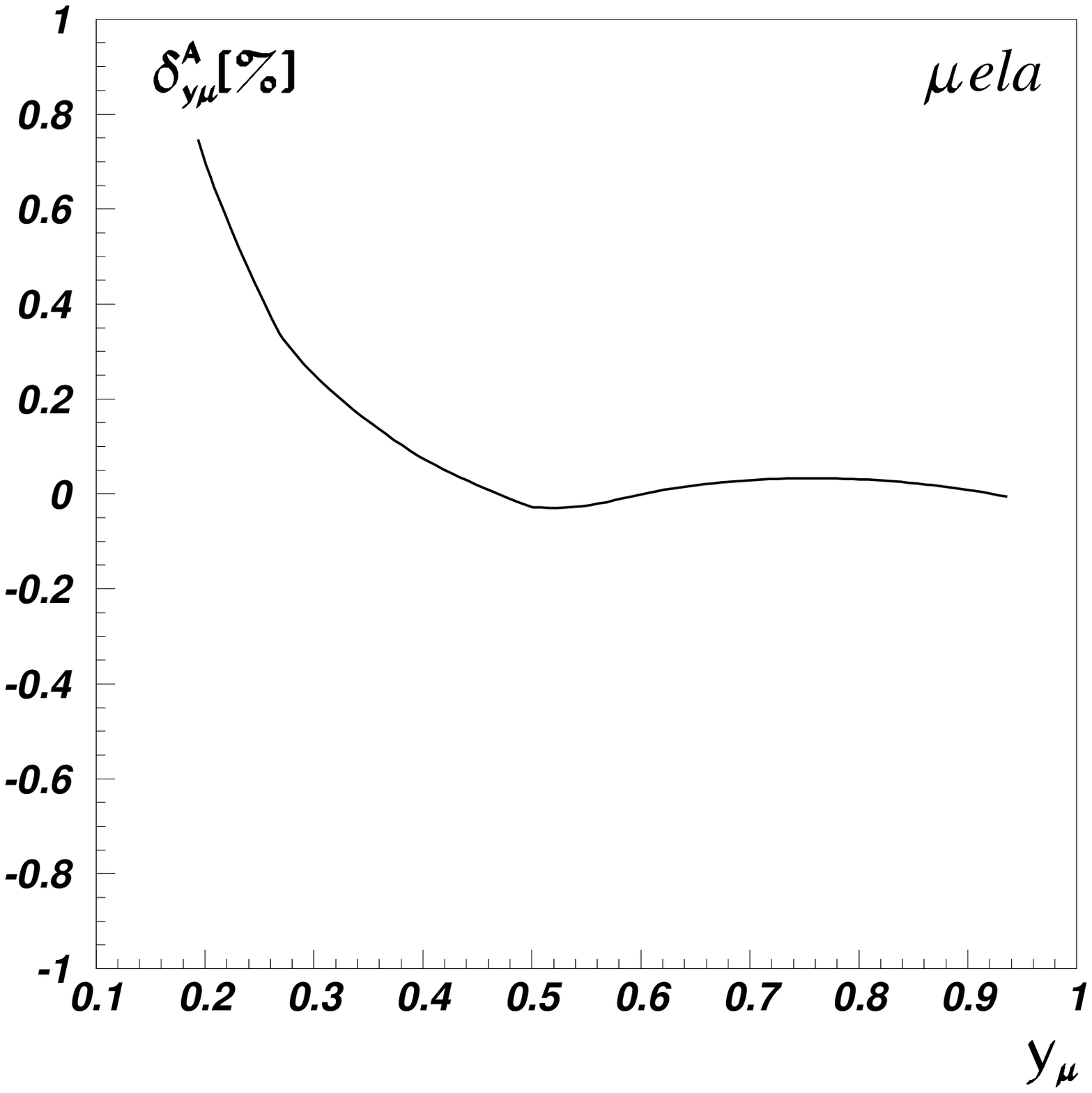,bbllx=0pt,bblly=0pt,%
bburx=600pt,bbury=600pt,width=10cm,angle=0}}
\end{center}
\vspace{-4.0cm}
\caption[xxx]{\sf The QED radiative corrections to asymmetry with
experimental cuts.}
\label{fiia2}
\end{figure}
\begin{figure}[tb]
\center
\vspace{-8mm}
\mbox{\epsfig{file=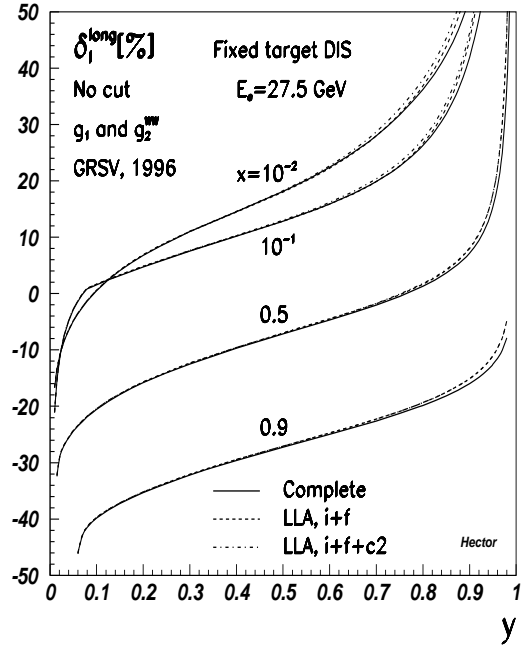,height=10cm,width=8cm}}
\vspace{-10mm}
\caption[xxx]{\sf
A comparison of complete and LLA RC's in the kinematic regime
of HERMES for neutral current longitudinally
polarized DIS in leptonic variables.
The polarized parton densities \cite{96} are used. The structure
function $g_2$ is calculated using the Wandzura--Wilczek relation.
$c2$ stands for the Compton contribution, see~\cite{ph} for details.}
\label{fig1}
\end{figure}
\begin{figure}[tb]
\center
\vspace{-5mm}
\mbox{\epsfig{file=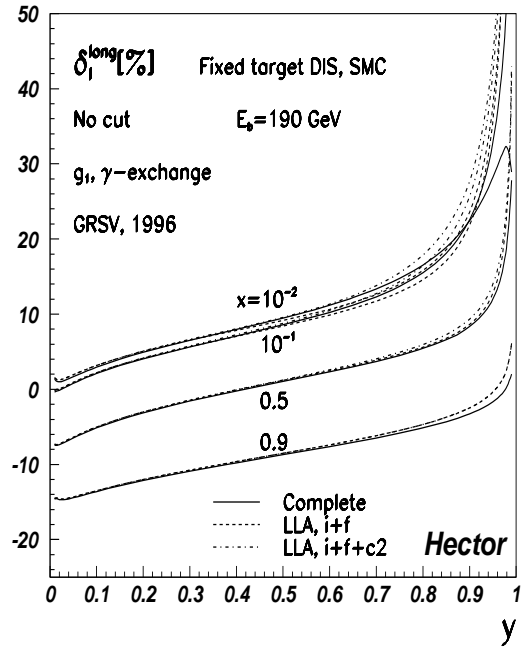,height=10cm,width=8cm}}
\vspace{-8mm}
\caption[xxx]{\sf
The same as in~fig.~\ref{fig1}, but for energies in the
range of the SMC-experiment.}
\label{fig2}
\end{figure}
\clearpage
\begin{figure}[tb]
\center
\vspace{-5mm}
\mbox{\epsfig{file=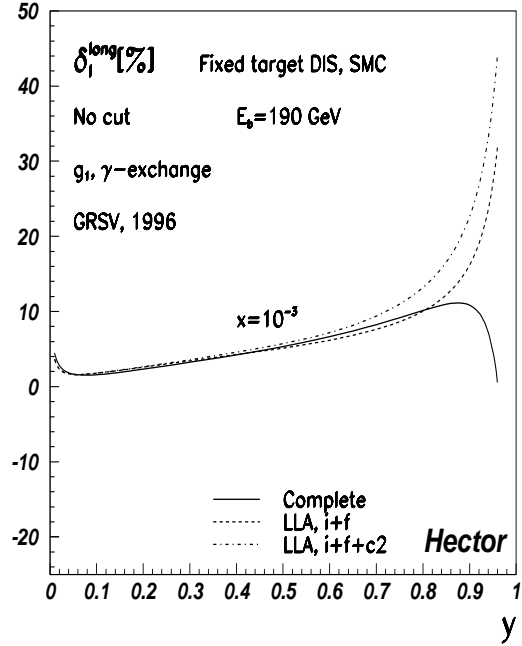,height=10cm,width=8cm}}
\vspace{-5mm}
\caption[xxx]{\sf
The same as in fig.~\ref{fig2} for $x=10^{-3}$.}
\label{fig3}
\end{figure}
\begin{figure}[tb]
\center
\vspace{-5mm}
\mbox{\epsfig{file=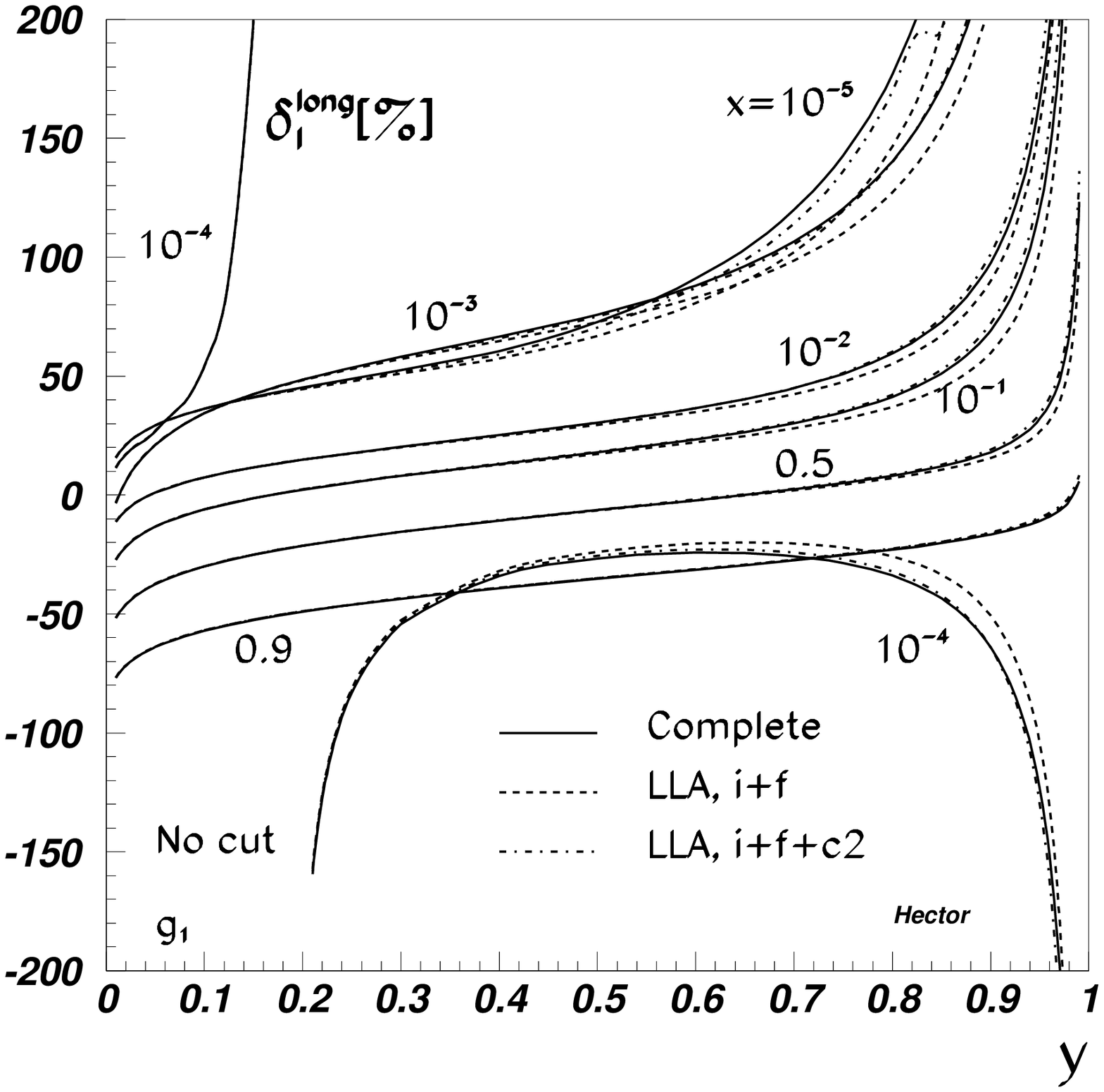,height=10cm,width=8cm}}
\vspace{-5mm}
\caption[xxx]{\sf
A comparison of complete and LLA RC's at HERA collider
kinematic regime
for neutral current deep inelastic scattering off a longitudinally
polarized target measuring the kinematic variables at the leptonic
vertex.}
\label{fig4}
\end{figure}

\begin{thebibliography}{99}
%
\bibitem{h91}
{\sf Proceedings of the
Workshop on Physics at HERA, 1991
Hamburg (DESY, Hamburg, 1992)}, W.~Buchm\"uller and G.~Ingelman (eds.).
%
\bibitem{JBKIN}
J.~Bl\"umlein, Z. Phys. {\bf C65} (1995) 293.
%
\bibitem{tp}
D.~Bardin, L.~Kalinovskaya and T.~Riemann,
DESY 96--213,   {Z. Phys.} {\bf C} in print.
%
\bibitem{muela}
D.~Bardin and L.~Kalinovskaya, $\mu${\bf e}{\it la}, version 1.00,
March 1996.
The source code is available from {\tt http://www.ifh.de/\~{}bardin}.
%
\bibitem{he}
A.~Arbuzov, D.~Bardin, J.~Bl\"umlein, L.~Kalinovskaya and T.~Riemann,
Comput. Phys. Commun. {\bf 94} (1996) 128, {\tt hep-ph/9510410}
%
\bibitem{ph}
D.~Bardin, J.~Bl\"umlein, P.~Christova and L.~Kalinovskaya,
DESY 96--189, {\tt hep-ph/9612435}, {Nucl. Phys.} {\bf B} in print.
%
\bibitem{SMC}
SMC collaboration,
D.~Adams et al., Phys. Lett. {\bf B396} (1997) 338;
Phys. Rev. {\bf D56} (1997) 5330, and references therein.
%
\bibitem{blkwasborn}
A.I.~Nikischov, Sov. J. Exp. Theor. Phys. Lett. {\bf 9} (1960) 757;\\
P. van Nieuwenhuizen, Nucl. Phys. {\bf B28} (1971) 429;\\
D.~Bardin and N.~Shumeiko, Nucl.Phys. {\bf B127} (1977) 242.
%
\bibitem{bla}
T.V.~Kukhto, N.M.~Shumeiko and S.I.~Timoshin,
J. Phys. {\bf G13} (1987) 725.
%
\bibitem{da}
G.~Passarino, {Comput. Phys. Commun.} {\bf 97} (1996) 261.
%
\bibitem{crad96}
D.~Bardin, J.~Bl\"umlein, P.~Christova, L.~Kalinovskaya, and T.Riemann,
Acta Phys. Polonica {\bf B28} (1997) 511.
%
\bibitem{bk1}
J~.Bl\"umlein and N.~Kochelev, {Phys. Lett.} {\bf B381} (1996) 296;
Nucl. Phys. {\bf B498} (1997) 285.
%
\bibitem{hy}
D.~Bardin, J.~Bl\"umlein, P.~Christova and L.~Kalinovskaya,
Preprint  DESY 96--198, {\tt hep-ph/9609399},
in: {\sf Proceedings of the
  Workshop  `Future Physics at HERA'}, G.~Ingelman, A.~De~Roeck,
  R.~Klanner (eds.), {\bf Vol.~1},~p.~13; {\tt hep-ph/9609399}.
%
\bibitem{db}
D.~Bardin, Contribution to the Proceedings of the
International Conference
on High Energy Physics, Warsaw, August 1996.
%
\bibitem{96}
M.~Gl\"uck, E.~Reya, M.~Stratmann and W.~Vogelsang,
{Phys. Rev.} {\bf D53} (1996) 4775.
%
\bibitem{ww}
S.~Wandzura and F.~Wilczek,
{Phys. Lett.} {\bf B72} (1977) 195.
%
\bibitem{pr}
I.Akushevich, A.Il'ichev, N.Shumeiko, A.Soroko and A.Tolkachev,
{\tt hep-ph/9706516}.
\end{thebibliography}
\end{document}